\documentclass[preprint,showpacs,preprintnumbers,amsmath,amssymb,superscriptaddress]{revtex4}

\usepackage{graphicx}
\usepackage{dcolumn}
\usepackage{bm}

\begin{document}


\title{Klein-tunneling transistor with ballistic graphene}

\author{Quentin Wilmart}
\affiliation{Laboratoire Pierre Aigrain, Ecole Normale Sup\'erieure, CNRS (UMR 8551),
Universit\'e P. et M. Curie, Universit\'e D. Diderot,
24, rue Lhomond, 75231 Paris Cedex 05, France}
\author{Salim Berada}
\affiliation{Institute of Fundamental Electronics, Univ. Paris-Sud, CNRS, Orsay, France}
\author{David Torrin}
\affiliation{D\'epartement de Physique, Ecole Polytechnique, 91128 Palaiseau, France}
\author{V. Hung Nguyen}
\affiliation{Institute of Fundamental Electronics, Univ. Paris-Sud, CNRS, Orsay, France}
\author{Gwendal F\`eve}
\affiliation{Laboratoire Pierre Aigrain, Ecole Normale Sup\'erieure, CNRS (UMR 8551),
Universit\'e P. et M. Curie, Universit\'e D. Diderot,
24, rue Lhomond, 75231 Paris Cedex 05, France}
\author{Jean-Marc Berroir}
\affiliation{Laboratoire Pierre Aigrain, Ecole Normale Sup\'erieure, CNRS (UMR 8551),
Universit\'e P. et M. Curie, Universit\'e D. Diderot,
24, rue Lhomond, 75231 Paris Cedex 05, France}
\author{Philippe Dollfus}
\affiliation{Institute of Fundamental Electronics, Univ. Paris-Sud, CNRS, Orsay, France}
\author{Bernard Pla\c{c}ais}
\email{Bernard.Placais@lpa.ens.fr}
\affiliation{Laboratoire Pierre Aigrain, Ecole Normale Sup\'erieure, CNRS (UMR 8551),
Universit\'e P. et M. Curie, Universit\'e D. Diderot,
24, rue Lhomond, 75231 Paris Cedex 05, France}

\date{\today}

\begin{abstract}
Today the availability of high mobility graphene up to room temperature makes ballistic transport in nanodevices achievable.
In particular, p-n-p transistor in the ballistic regime gives access to the Klein tunneling physics and allows the realization of devices exploiting the optics-like behavior of Dirac Fermions (DF) as in the Vesalego lens or the Fabry P\'erot cavity.
Here we propose a Klein tunneling transistor based on geometrical optics of DF. We consider the case of a prismatic active region delimited by a triangular gate, where total internal reflection may occur, which leads to the tunable suppression of the transistor transmission. We calculate the transmission and the current by means of scattering theory and the finite bias properties using Non Equilibrium Green's Function  (NEGF) simulation.

\end{abstract}

\maketitle

\section{Introduction}

As a high mobility material, graphene is well suited for high frequency electronics \cite{Novoselov2012Nature,Wu2012aNanoL}. Additionally, thanks to the weakness of electron-phonon coupling \cite{Chen2008nnano,Efetov2010PRL,Betz2012PRL,Betz2013nphys}, the high mobility persists at high temperature \cite{Wang2013Science} so that one can envision ballistic graphene electronics in realistic operating conditions. The absence of band-gap restricts the switching off capability of graphene, the conductivity reaching a shallow minimum at charge neutrality limited by the quantum of conductance $4e^2/h$. For logic electronic applications a band-gap can be restored using \emph{e.g.} geometrical confinement \cite{Meng2011ieee}, but this is usually at the cost of a reduced mobility. In microwave electronics both a high mobility and a significant switching capability are needed to achieve large voltage and power gain at high frequency.  To this end we explore here the new possibilities offered by Klein tunneling (KT), a hallmark of Dirac fermions (DFs).

We consider the case of p-n-p transistors where the resistance at the source-gate and gate-drain junctions is dominated by the \emph{Klein tunneling effect}, while DFs propagate ballistically in the barrier \cite{Katnelson2006nphys}. Graphene being ambipolar the generalization to n-p-n transistors is straightforward.
This is the regime of DF optics where refraction and transmission at the interfaces are determined by Fresnel-like relations \cite{Allain2011EPJB} and depend on the abruptness of junctions \cite{Cheianov2007Science,Cayssol2009PRL}. Architectures exploiting optical properties of DFs have been already proposed like the Vasalego lens \cite{Cheianov2007Science,Williams2011nnano}, and demonstrated like Fabry-P\'erot interferometers \cite{Young2009NPHYS,Wu2012bNanoL,Rickhaus2013ncom} or tilted p-n junctions \cite{Sajjad2012PRB,Sutar2012NanoL} (for a review, see Ref.\cite{Rozhkov2011PhysRep}). The possibility of a "latticetronics" of Klein tunneling currents in an armchair ribbon using potential barriers has also been recently discussed \cite{Lopez2014JPCM}. Here we consider a different geometry, that exploits total internal reflection in a \emph{Klein tunneling prism}. The prism is made of an n-doped ballistic triangular domain embedded in a p-doped diffusive area, the latter being controlled electrostatically or chemically. Alike in light reflectors, an array of KT-prisms can be used to form the active channel of a \emph{Klein tunneling transistor}; this geometry minimizes the gate length to keep ballistic transport conditions in a way similar to Fresnel lenses which minimize glass weight and light absorption \cite{Fresnel1875book}. In this work we calculate the low-energy transmission of such a device using scattering theory, i.e. within an intuitive and physically transparent approach. Our study is supplemented by atomistic simulations using the non-equilibrium Greens function (NEGF) formalism that includes a more complete description of quantum transport. It gives access to finite bias properties while accounting for short gate effects like direct drain-source tunneling, diffraction, dispersion and finite temperature effects.
Both approaches predict for the Klein tunneling transistor a strong suppression of conductance that can eventually go below the minimum conductance at charge neutrality. This transistor can be used as a tunable barrier for electrostatic quantum confinement to achieve, e.g. single Dirac fermion pumps working at low temperature. It is also suited for microwave electronics as it cumulates significant resistance in the OFF state with a large conductance in the ON state. A nanoscale variant has been recently proposed that predicts very large
ON-OFF ratios for logic applications \cite{Jang2013pnas}.
Our approach is different;  it is based on geometrical DF optics following Ref.\cite{Torrin2010Report}; it is more conservative in accounting for diffusive transport in the leads and targets microwave electronic applications \cite{Schwierz2010nnano, Pallecchi2011APL}.

\begin{figure}[ttt]
\includegraphics[scale=0.3]{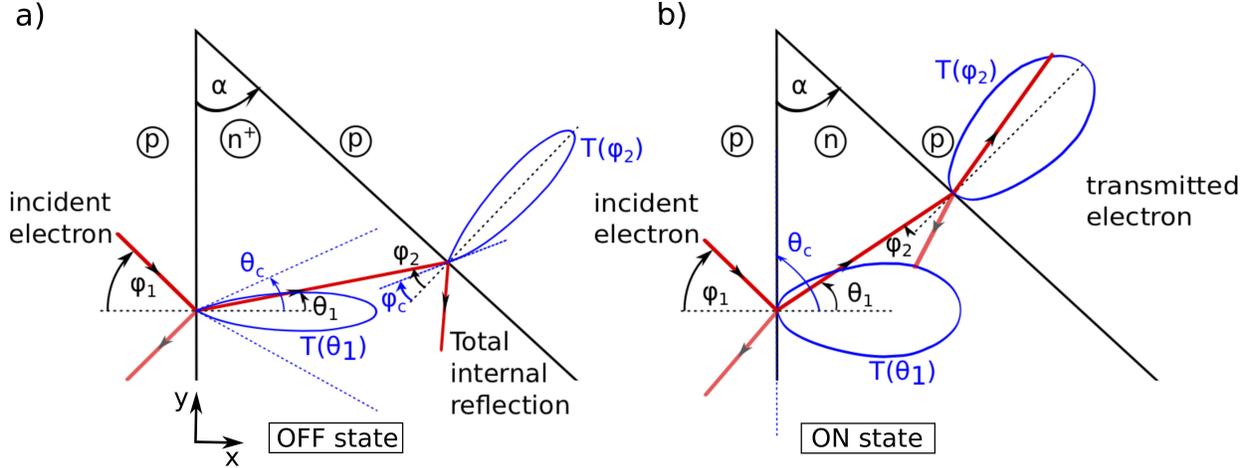}
\caption{Principle of total reflection in a Klein tunneling prism. The refraction angle of Dirac fermion (DF) beams (red rays) and their angular dependent transmission amplitude (blue lobes) are controlled by the optical-like index ratio $\nu=-\sqrt{n/p}$ of the p and n regions.
a) OFF state $n\gg p$. Anisotropic forward scattering occurs at p-n junction : the refracted rays are mostly transmitted along the junction normal within a lobe limited by $|\theta_1| <\theta_c$.
The n-p  junction selects the incident carriers that are close to the normal to the junction (i.e. $|\phi_2 |< \phi_c $); other being reflected.
b) ON state $n\simeq p$. In this case $\theta_c=\phi_c=90\deg$ which means all indent rays are transmitted with large transmission coefficient at both interfaces.}
\label{Willmart_prism}
\end{figure}

The principle of the Klein tunneling prism is sketched in figure \ref{Willmart_prism}. It can be understood by scattering theory and  relies on the total internal reflection in a triangular n-doped graphene region (concentration $n$,  angle $\alpha$), the refractive medium, embedded in a p-doped environment (concentration $p\lesssim n$) playing the role of vacuum. The refraction at the input p-n junction obeys a Snell-Descartes-like relation  $\sin\phi_1=\nu\sin\theta_1$ where $\nu=-\sqrt{n/p}$ is the (negative) refraction index \cite{Allain2011EPJB} as sketched in the figure. The transmission $\cal{T}(\phi)$ can be calculated for a sharp junction\cite{Cheianov2007Science}, but also for smooth junctions (see below and Ref.\cite{Cayssol2009PRL}). The latter case is more realistic and suitable for device modeling. The main feature of Klein tunneling is the enhanced forward scattering for $n\gg p$. As shown in the figure, the refracted DF beam is focussed along the junction normal within an angular opening  $|\theta_1|\leq\theta_c$, where  $\theta_c=\arcsin{1/\nu}$ is the critical angle.  At the drain side, which is an n-p junction where DF are impinging at incidence $\phi_2=\alpha-\theta_1$, the Snell-Descartes relation reads $\nu\sin\phi_2=\sin\theta_2$.  For $|\nu|\gg1$, this implies total reflection for $|\phi_2|\geq\phi_c$, with $\phi_c=\theta_c$ for symmetric drain and source doping (and zero bias).
The condition is met for any ray incident to the prism provided that $\theta_c<\alpha/2$ or $n>p/(sin{(\alpha/2)})^2$. Finally, the reflected beams are transmitted back across the source junction. The reflection can be controlled, and transmission restored, on decreasing $n$-doping, i.e. increasing $\theta_c$, as shown in figure \ref{Willmart_prism}-b. Note that one obtains an equivalent effect on decreasing the prism angle down to $\alpha=0$; it means that the transmission in the open state $n\simeq p$  (or $\nu=-1$) should approach that of a rectangular gate transistor. Taking $\alpha=45\deg$, one can estimate the gate doping for full reflection : $n>6.8 p$ (or $|\nu|>2.6$). The transmission of a KT transistor is suppressed here when increasing the gate doping deep in the metallic regime; this is a marked difference with  conventional semiconducting transistors  where transmission is suppressed on pinching-off the channel.
Operating a transistor  with a large intrinsic transmission should benefit for the dynamical and noise properties of the device.
In order to establish the above sketch, a full calculation is needed that accounts for finite transmission and multiple reflection effects.

\begin{figure}[ttt]
 \includegraphics[scale=0.4]{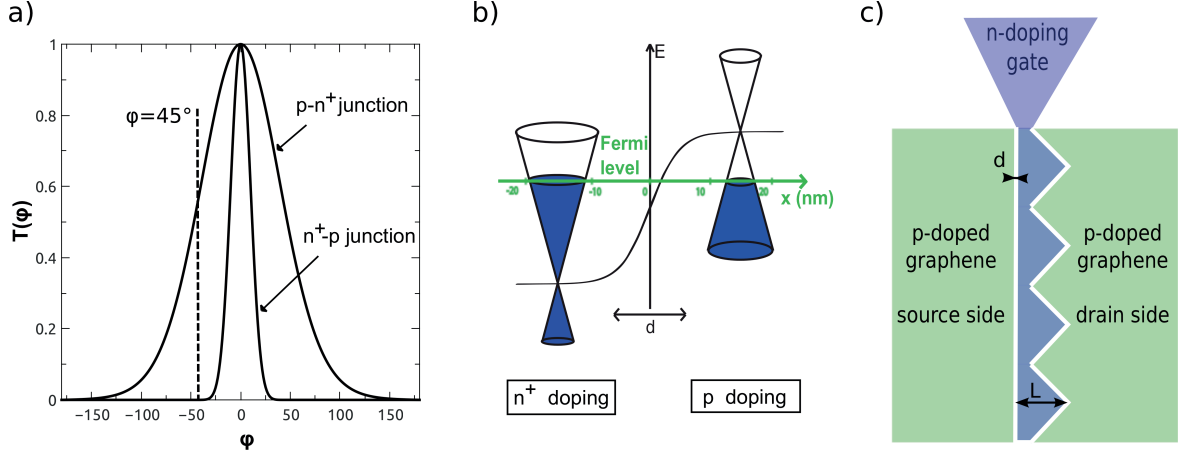}
\caption{ a) Incident angle $\phi$ dependence of the transmission coefficient for smooth junction ($d=10\;\mathrm{nm}$). Angular dependent transmission at a $p-n^+$ and a $n^+-p$ junctions ($n^+=6 p$); the contrast is most pronounced for an incidence angle  $\phi=45\deg$.
b) Sketch of the Fermi function potential profile and associated band structure of a $n^+-p$ junction with a length of the junction $d=10\;\mathrm{nm}$.
c) Sketch of a Klein tunneling transistor with its split gates. The barrier area is blue (n-doping gate) and the leads areas are green (p-doping gate). Here the Klein tunneling transistor is made of four elementary units.}
\label{collimation}
\end{figure}

We rely on a realistic modeling of the p-n junctions. Translational invariance along the junction parallel to $y$-axis yields the above mentioned Snell-Descartes relation between $\theta$ and $\phi$.
The transmission of a smooth junction, characterized by a
length $d\gg k_p^{-1}$ where $k_p$ (resp. $k_n=|\nu|k_p$) is the Fermi wave number in p-doped (resp. n-doped) region, has a  transmission ${\cal T}_{smooth}(\phi)\simeq \exp{-[\pi d k_p^2\sin{\phi}^2/(k_p+k_n)]}$
\cite{Cheianov2007Science}.
Dealing with  intermediate junctions length  ($d\sim k_p^{-1}$), we rely on the expression by Cayssol et al. Ref.\cite{Cayssol2009PRL}, which is the exact solution for a potential step described by a Fermi function, $V(x)=V_0(1+\exp{(-x/w)})^{-1}$ (see figure \ref{collimation}-b).
We use $d=4.5w$ to make contact with the smooth junction formula.
The transmission is given by ${\cal T}(\phi)= 1-[\sinh{(\pi w k^{+-})}\sinh{(\pi w k^{-+})}]/[\sinh{(\pi w k^{++})}\sinh{(\pi w k^{--})}]$,
where $k^{\alpha\beta}=k_p(1+\alpha\cos{\phi})+k_{n}(1+\beta \cos{\theta})$. The angular dependence ${\cal T}(\phi)$ is displayed in Fig.\ref{collimation}-a for two representative configurations of the OFF state:   p-n$^+$ and n$^+$-p junctions with $n^+=6p$. The contrast is most pronounced for $\phi\gtrsim45\deg$ with $T_{p-n^+}\lesssim 0.6$ and $T_{p-n^+-p}=0$, which is the regime of total internal reflection. The contrast is close to that predicted for a sharp junction but different from that of smooth junction (not shown) which justifies the use of the Cayssol interpolation formula.

\begin{figure}[ttt]
 \includegraphics[scale=0.5]{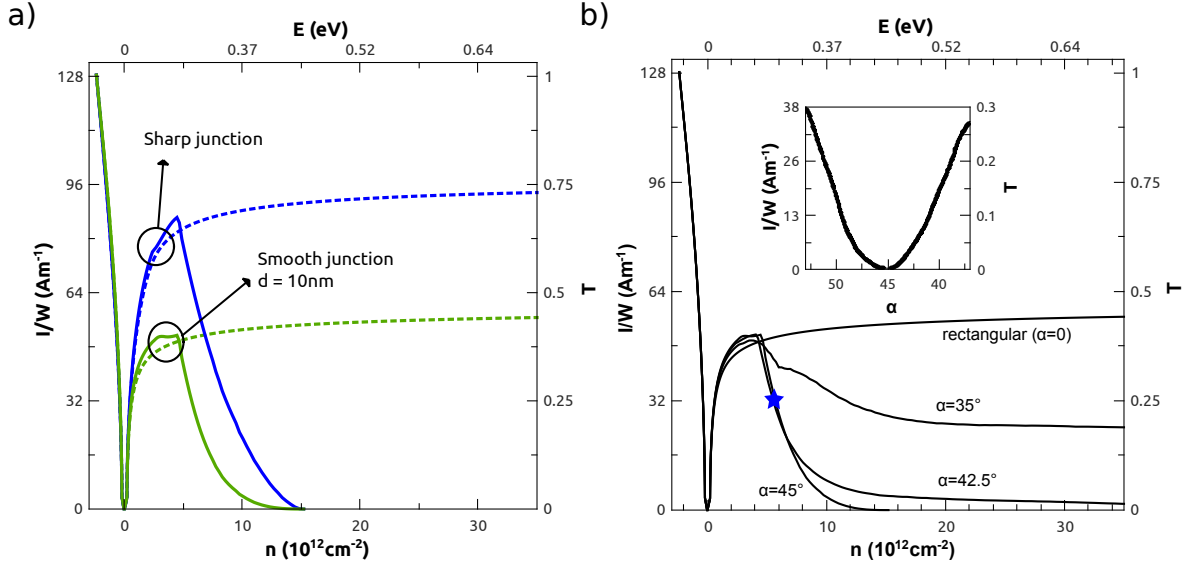}
\caption{Current density $I/W$ (and corresponding transmission $T$) from scattering theory of various Klein tunneling p-n-p transistors as a function of gate doping $n$; with leads doping of $p=2.3\times10^{12}\;\mathrm{cm^{-2}}$, for rectangular gate (dotted lines in panel a) and saw-tooth gate (solid lines). The rectangular gate reproduces the well known graphene transistor behaviour while the triangular geometry has its transmission suppressed at high doping.
a) Sharp junction (blue lines) shows higher transmission than smooth junction (green lines) with junction length of $d=10\;\mathrm{nm}$.
b) Transmission for various gate geometry from rectangular ($\alpha \xrightarrow{} 0$) to $\alpha=45\deg$. Inset : $\alpha$ dependence of current density (and corresponding transmission) at $n=6p$ with a minimum for the $\alpha=45\deg$ geometry.
The star indicates the working point of a rf transistor.} \label{KT_scattering_model}
\end{figure}

The principle of the Klein tunneling prism can be realized in a Klein tunneling transistor (figure \ref{collimation}-c) on stacking, head-to-tail, a series of prims to realize a sawtooth gate with symmetric elementary units made of an isosceles triangle of opening angle $\pi-2\alpha$. This geometry allows to implement the refractor  principle while keeping the gate length short enough to remain in the ballistic regime. The n-doped barrier is assumed to be ballistic and controlled by the transistor gate. The leads can be either access regions
or simply metallic contacts.  In the latter case, carrier concentration is set by electro-chemical doping, in the former it can be tuned by a second electrostatic gate. To assess device properties, we calculate the transmission of DF rays across an elementary triangle. We consider a beam with a given incidence angle $\phi_1$ and position $y$
along the source side. The trajectory of the ray into the prism is calculated according to above p-n junction refraction and reflection rules iterated up to twenty
internal reflections.  The beam intensity inside the prism decreases at each reflection,  the transmitted intensity outside the prism increases accordingly. We then integrate over $y$ positions and all incident angles (diffusive leads) weighted by the angular
density of states to get the overall transmission $T=\langle{\cal T}\rangle_{\phi,y}$.
Results are plotted in Fig.\ref{KT_scattering_model} as function of gate doping, for $p=2.3\times 10^{12}\;\mathrm{cm^{-2}}$.
We estimate the current density per unit width  $W$,  $I/W=(4e^2/h)(k_p/\pi)TV_{ds}$, which is  plotted here for a small $V_{ds}=10\;\mathrm{mV}$ bias.
Panel a) shows the effect of the junction sharpness for both rectangular and triangular gates; panel b) that of the prism angle keeping the gate area constant.
In the standard Klein tunneling with rectangular gates, the transmission saturates at $75\%$ (resp. $45\%$) for sharp (resp. $10\;\mathrm{nm}$-long)
junctions which determines the ON-state current and conductance $g_{ON}/W\simeq 5\;\mathrm{mS \mu m^{-1}}$  ($10\;\mathrm{nm}$-long junction).
As seen in panel b), the OFF-state current is very sensitive to the prism opening angle; it is zero for $\alpha=45\deg$ (and $n\gtrsim 6p$) but increases
rapidly whenever $\alpha$ deviates by more than $5\deg$  from this value (see inset of figure \ref{KT_scattering_model}-b).
From this analysis, we conclude that the optimal geometry for a KT transistor is $\alpha=45\pm5\deg$ and abrupt junctions
($d\lesssim10\;\mathrm{nm}$).
This estimate is characteristic of the robustness of the device against geometrical imperfections, like prism asymmetry, apex rounding or p-n junction roughness.
In these conditions, one obtains a significant modulation of transmission in the range $0$--$0.4$ which is
appropriate for electrostatic confinement in quantum dots.
From the slope of the $I(V_{ds})$ curve
(working point labeled by a star in Figure \ref{KT_scattering_model}-b) and considering a gate capacitance $c_g=10\;\mathrm{fF\mu m^{-2}}$ (equivalent SiO$_2$ thickness of 3 nm), we estimate the bias dependence of transconductance
per unit width $g_mV_{ds}^{-1}/W\simeq8\;\mathrm{mS\mu m^{-1}V^{-1}}$, a voltage gain dependence $A_VV_{ds}^{-1}=g_mV_{ds}^{-1}/g_{ds}\gtrsim3\;\mathrm{V^{-1}}$
and a transit frequency $f_T/V_{ds}=g_m/2\pi c_gL\simeq 1300\;\mathrm{GHz/V}$ (for a gate length $L=100\;\mathrm{nm}$).

We discuss now  the physical limits of our geometrical optics description which relies on the assumption that $d\lesssim k_p^{-1}\ll L$. The refraction principle is scale independent but in practice the characteristic length of the gate $L$ should be smaller than the ballistic length, while remaining larger than electronic wave length to avoid deleterious effects of diffraction.  The principle of the reflector being very sensitive to refraction angles (see inset of Fig.\ref{collimation}), an uncertainty  $\Delta\theta\sim (k_pL)^{-1}$ in angular orientation, due to finite wavelength, may drastically affect the total internal reflection.   Another limitation for the OFF state resistance is the direct drain source tunneling which is also strongly scale dependent. Finally we have neglected for simplicity dispersion effects arising from broad sources such as finite bias DF emission. In order to quantify these effects we have performed a numerical simulation of the KT transistor by means of NEGF method. Our model is based on a tight-binding Hamiltonian to describe the electron state in the graphene honeycomb lattice \cite{Berrada2013APL}.

\begin{figure}[ttt]
\includegraphics[scale=0.6]{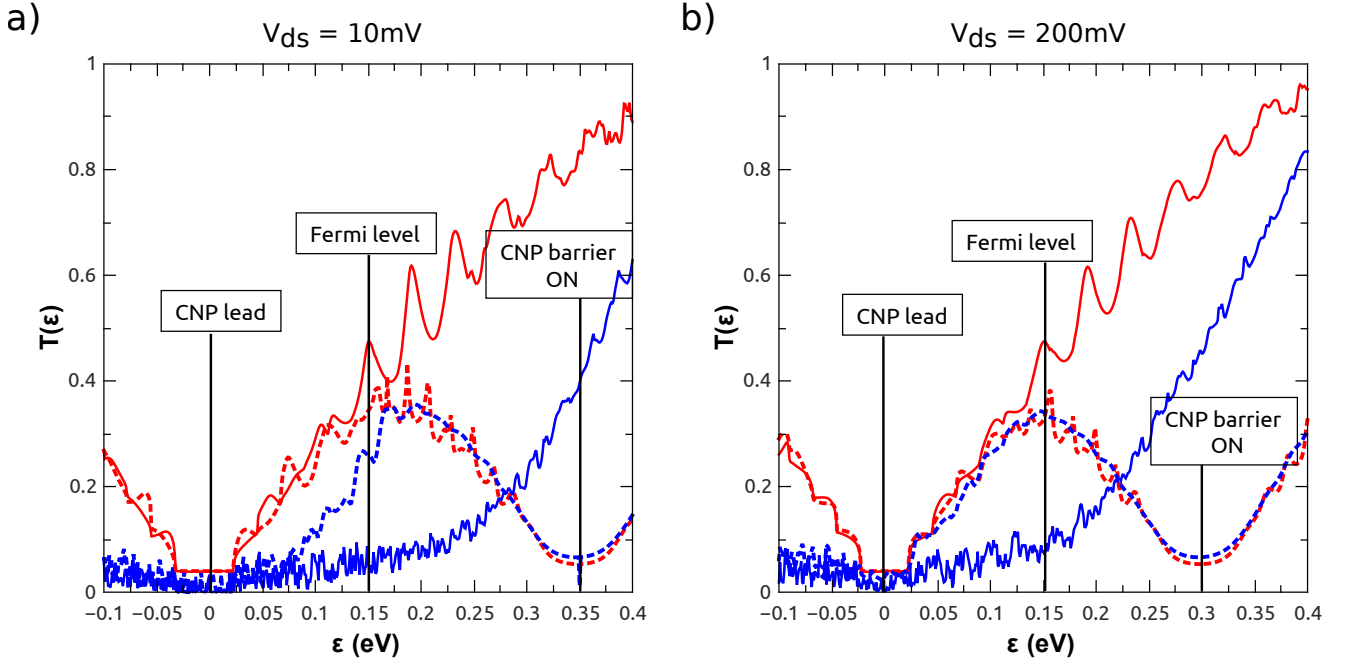}
\caption{ Transmission coefficient of the device from the NEGF simulations as a function of carrier energy $\epsilon$ for a given barrier height (from gate doping $n$) and a given Fermi level (from leads doping $p$). We choose $\epsilon=0$ at the CNP of the lead. Dotted lines (resp. solid lines) correspond to the ON state $n\sim 4\times 10^{12}\;\mathrm{cm^{-2}}$ (resp. OFF state $n\sim 6\times 10^{13}\;\mathrm{cm^{-2}}$ ). Red lines (resp. blue lines) correspond to rectangular (resp. triangular) device. Both panel a (low bias $V_{ds}=10mV$) and panel b (high bias $V_{ds}=200mV$) shows a similar transmission around the Fermi level in the OFF state for rectangular and triangular devices while transmission is suppressed in the OFF state for triangular device only.} \label{transmission}
\end{figure}

\begin{figure}[ttt]
\includegraphics[scale=0.5]{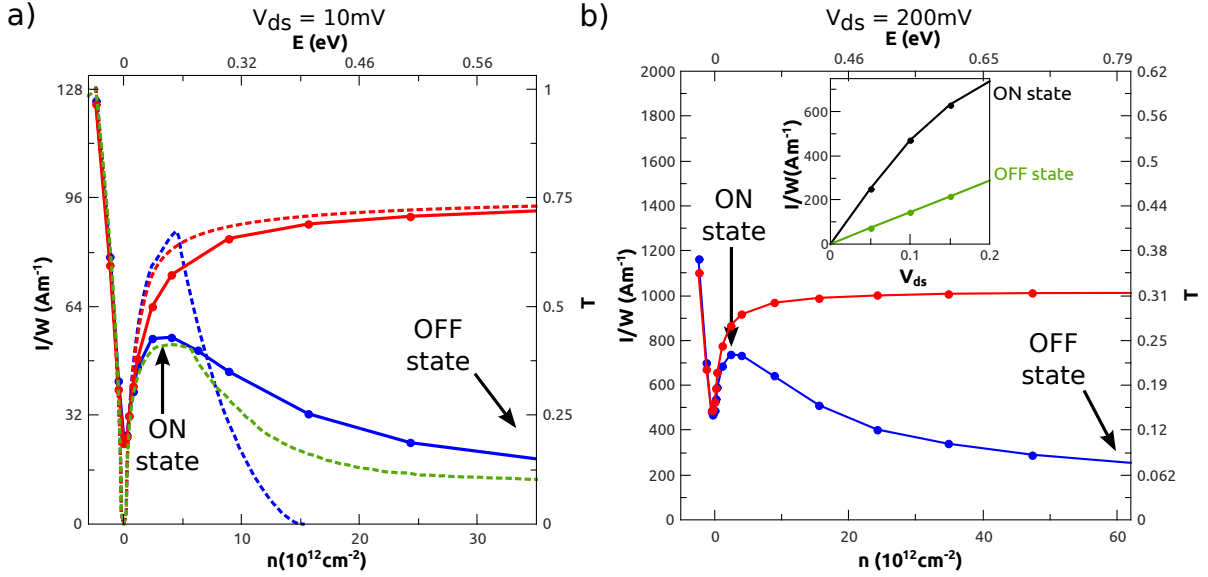}
\caption{ Barrier doping $n$ dependence of current density $I/W$ (and corresponding total transmission $T$) for a) $V_{ds}=10\;\mathrm{mV}$ and b) $V_{ds}=200\;\mathrm{mV}$. Solid lines (resp dotted lines) correspond to NEGF simulations (resp. scattering calculations), red lines (resp. blue lines) correspond to rectangular (resp. triangular) device. In panel a) scattering calculation for rectangular device with sharp junctions (red dotted line) is in good agreement with NEGF. Scattering calculation for triangular device with smooth junction $d=10\;\mathrm{nm}$ and opening angle $\alpha=52\deg$ (green dotted line) is a way to take into account diffraction effects and compares well with NEGF simulation (blue solid line).
Panel b) inset : $I-V_{ds}$ characteristics of the triangular transistor in the ON state $n= 2\times 10^{12}\;\mathrm{cm^{-2}}$ (black line) and in the OFF state $n=6\times 10^{13}\;\mathrm{cm^{-2}}$ (green line)} \label{FIG5_NEGF}
\end{figure}

NEGF simulation assumes a transistor of ballistic graphene whose size is scaled down by a factor 2.5 with respect to a realistic device to make the calculation achievable with available computational resources. Additionally, the self-consistent solution of Poisson's equation has been deactivated here to save CPU time. The gate-induced potential in graphene was modelled as a "square" barrier of height reduced by $V_{ds}/2$ at the source-end and enhanced by $V_{ds}/2$ at the drain-end. This simple description of potential has been shown realistic from self-consistent simulation of GFETs in the p-n-p or n-p-n regimes \cite{Alarcon2013IEEE}.
The transistor gate is taken either rectangular or triangular but with the same gate area. The triangular gate is an elementary unit of Fig.\ref{collimation}-c. The channel width is $80\;\mathrm{nm}$, the gate length is $40\;\mathrm{nm}$ for the rectangular device. It is modulated between $20\;\mathrm{nm}$ and $60\;\mathrm{nm}$ for the triangular device. We set the source side doping to $p=2.3\times 10^{12}\;\mathrm{cm^{-2}}$ which corresponds to a Fermi energy $E_F=\hbar v_F\sqrt{\pi p}\simeq 0.15eV$. The NEGF simulation provides the transmission coefficient $T(\epsilon)$ through the device, where $\epsilon$ is the carrier energy, considering the charge neutrality point (CNP) of the source as the zero energy reference. Two situations are considered : a low bias case $V_{ds}=10\;\mathrm{mV}\ll E_F/e$ (Fig.\ref{transmission}-a and \ref{FIG5_NEGF}-a) and high bias case  $V_{ds}=200\;\mathrm{mV}\gtrsim E_F/e$ (Fig.\ref{transmission}-b and \ref{FIG5_NEGF}-b).
The former allows direct comparison with (zero bias) scattering calculations; the latter is typical of a transistor working point.

In Fig.\ref{transmission} $T(\epsilon)$ is displayed for the rectangular geometry (red lines) and the triangular geometry (blue lines) for the ON state of the transistor (dotted lines) and the OFF state (solid lines). In the ON state (small gate doping) two minima of transmission appear corresponding to the energy position of the CNP in the lead and in the barrier. A transmission maximum is observed that lies mid-way between both CNP's.  As $n\sim 2 p$, the maximum coincides with the Fermi energy, irrespective of the gate geometry. The transmission is large as expected for an ON state. In the OFF state (large gate doping) the transmission maximum is shifted toward higher energy and scaled up. The transmission at the Fermi level becomes sensitive to the gate geometry : it is large for rectangular gate and small for the triangular one. Oscillation of $T(\epsilon)$ observed for rectangular gates reflect Fabry P\'erot interferences; they are blurred for triangular gates. This effect has been neglected in the geometrical optics approach.

From the transmission, one calculates the current flowing through the device on integrating transmission in the Fermi energy interval between source and drain.
Finally we plot in Fig.\ref{FIG5_NEGF} the current density $I/W$ (and corresponding transmission $T$) as a function of gate doping $n$. Rectangular device transmission (red solid lines) and triangular device transmission (blue solid lines) from NEGF simulations are given for both low bias ($V_{ds}=10\;\mathrm{mV}$ Fig. \ref{FIG5_NEGF}-a) and high bias ($V_{ds}=200\;\mathrm{mV}$ Fig. \ref{FIG5_NEGF}-b). To compare with the scattering theory results we also plot in \ref{FIG5_NEGF}-a the sharp junction rectangular (resp. triangular) device transmission (red dotted line, resp. blue dotted line) from Fig.\ref{KT_scattering_model}-a.

First of all we notice that NEGF simulation for rectangular device is in close agreement with scattering theory. Regarding the triangular device, the agreement is more qualitative. The maximum current in the ON state is smaller and the OFF state current larger than geometrical optics prediction. We assign the difference to the diffraction effect that is more pronounced in small structures for a given Fermi wave length.  For instance, taking $k_n^{-1} \sim 10\;\mathrm{nm}$ gives an uncertainty in the prism geometry, in particular the opening angle with $\Delta\alpha\sim (kL)^{-1} \sim 15\% $. As an illustration, we have plotted in \ref{FIG5_NEGF}-a, the current for a triangular device with an opening angle of $1.15\alpha$ that compares favorably with the numerical calculation.
The high bias response in Fig. \ref{FIG5_NEGF}-b shows a comparable behavior. In both cases, numerical simulations show a current suppression below CNP minimum which is the signature of the Klein tunneling transistor. Our models correspond to two extreme cases of a macroscopic Klein tunneling transistor on one side and a nanometric device on the other one. An actual Klein tunneling transistor, with a gate length $L\lesssim 100\;\mathrm{nm}$ limited by the elastic mean free path, should have transport properties bracketed by this two limits. From a technical point of view, abrupt junctions can be realized using  local back gates, splitted by a nanometer thin gap and insulated from graphene by atomically thin hexagonal boron nitride layers \cite{Wang2013Science}.

In conclusion, we have introduced a Klein tunneling transistor architecture that takes advantage of anomalous refraction properties of Dirac Fermions in graphene to realize a tunable electrostatic barrier for Dirac fermions. We have used a geometrical optics model to explain transistor's principle. We have performed extensive numerical simulations of nanoscale variant of the device that confirm the transistor effect while taking a full account of finite size effects, in particular Dirac fermion diffraction. Our modeling will prove useful for the design of actual devices and to evaluate their potential in terms of single electron pumps and microwave electronics.

\begin{acknowledgments}
We thank Benjamin Huard for signaling us the interpolation formula for the junction collimation effect. This work was financially supported by the European Commission under the project Flagship-Graphene (contract no. 604391). The research has also been supported by the ANR-2010-MIGRAQUEL. The work of Q. Wilmart is supported by the French DGA.

\end{acknowledgments}

\end{document}